\documentclass[twocolumn,secnumarabic,amssymb, nobibnotes, aps, prd]{revtex4-1}

\usepackage{amsmath}
\usepackage{amssymb}
\usepackage{graphicx}
\begin{document}

\title{The effect of entanglement in gravitational photon-photon scattering}%

\author{Dennis R\"atzel, Martin Wilkens, Ralf Menzel}
\address{University of Potsdam, Institute for Physics and Astronomy\\
Karl-Liebknecht-Str. 24/25, 14476 Potsdam, Germany}

\begin{abstract}
The differential cross section for gravitational photon-photon scattering calculated in Perturbative Quantum Gravity is shown to depend on the degree of polarization entanglement of the two photons. The interaction between photons in the symmetric Bell state is stronger than between not entangled photons. In contrast, the interaction between photons in the anti-symmetric Bell state is weaker than between not entangled photons. The results are interpreted in terms of quantum interference, and it is shown how they fit into the idea of distance-dependent forces. 

\end{abstract}


\maketitle

\section{Introduction} Entanglement is an inherently quantum mechanical property. It is the basis of tests of non-classicality such as the renowned Bell-tests \cite{Bell1964,Aspect1982}. Hence, the effect of the entanglement of a physical system on its gravitational interactions  is in the overlap between quantum mechanics and gravity, which makes it a question of general physical interest.


In this article, we will investigate the gravitational effect of entanglement in PQG by calculating the differential cross section of gravitational photon-photon scattering for polarization entangled photons and not entangled photons. First, we shall shortly review the derivation of the polarization averaged differential cross section for gravitational photon-photon scattering in PQG. Then, we will consider polarization entangled photons. Finally, we will give an interpretation of the effect of entanglement on the differential cross section, firstly, from the perspective of quantum interference and, secondly, from the perspective of the localization of two-particle states.

\section{Photon-photon scattering} In Perturbative Quantum Gravity (PQG), the metric is written as $g_{\mu\nu}=\eta_{\mu\nu}+ \kappa h_{\mu\nu}$ and $h_{\mu\nu}$
becomes a quantum field of spin two.
The coupling to the electromagnetic field is given via the interaction Lagrangian
\begin{equation}
	\mathcal{L}_I:=-\zeta\,h_{\mu\nu}T^{\mu\nu}\,.
\end{equation}
with $\zeta:=\sqrt{8\pi G/c^3}$ \cite{Gupta1952,Boccaletti1967}.
It was shown by N. Grillo that $\mathcal{L}_I$ gives rise to a finite perturbation theory even in the first loop order (see remark and reference in \cite{Grillo2001scalar}). Here, only the first tree-level will be considered. PQG can be interpreted as a low energy effective theory of quantum gravity \cite{Donoghue2012}.

As a consequence of the peculiar split of the metric into an a priori background metric $\eta_{\mu\nu}$ and a quantum perturbation $h_{\mu\nu}$, we can avoid the conceptual problems of the standard interpretation of quantum mechanics applied to quantum gravity, i.e. that the classical observers doing preparations and measurements themselves live in the spacetime which they prepare and measure. Our classical observers live in the classical, flat spacetime given by $\eta_{\mu\nu}$. We assume that their detectors are separated by an infinite distance, and we assume that preparation and detection are performed at the temporal infinities. Furthermore, we assume that the gravitational interaction of the photons is the significant gravitational process beyond the effect of the classical metric background $\eta_{\mu\nu}$. Then, it seems safe to separate preparation and detection from the gravitational interaction process of the two photons, and the quantum properties of the metric only appear in our considerations via virtual gravitons mediating the interaction between the photons.
\begin{figure}[h]
\includegraphics[width=6cm,angle=0]{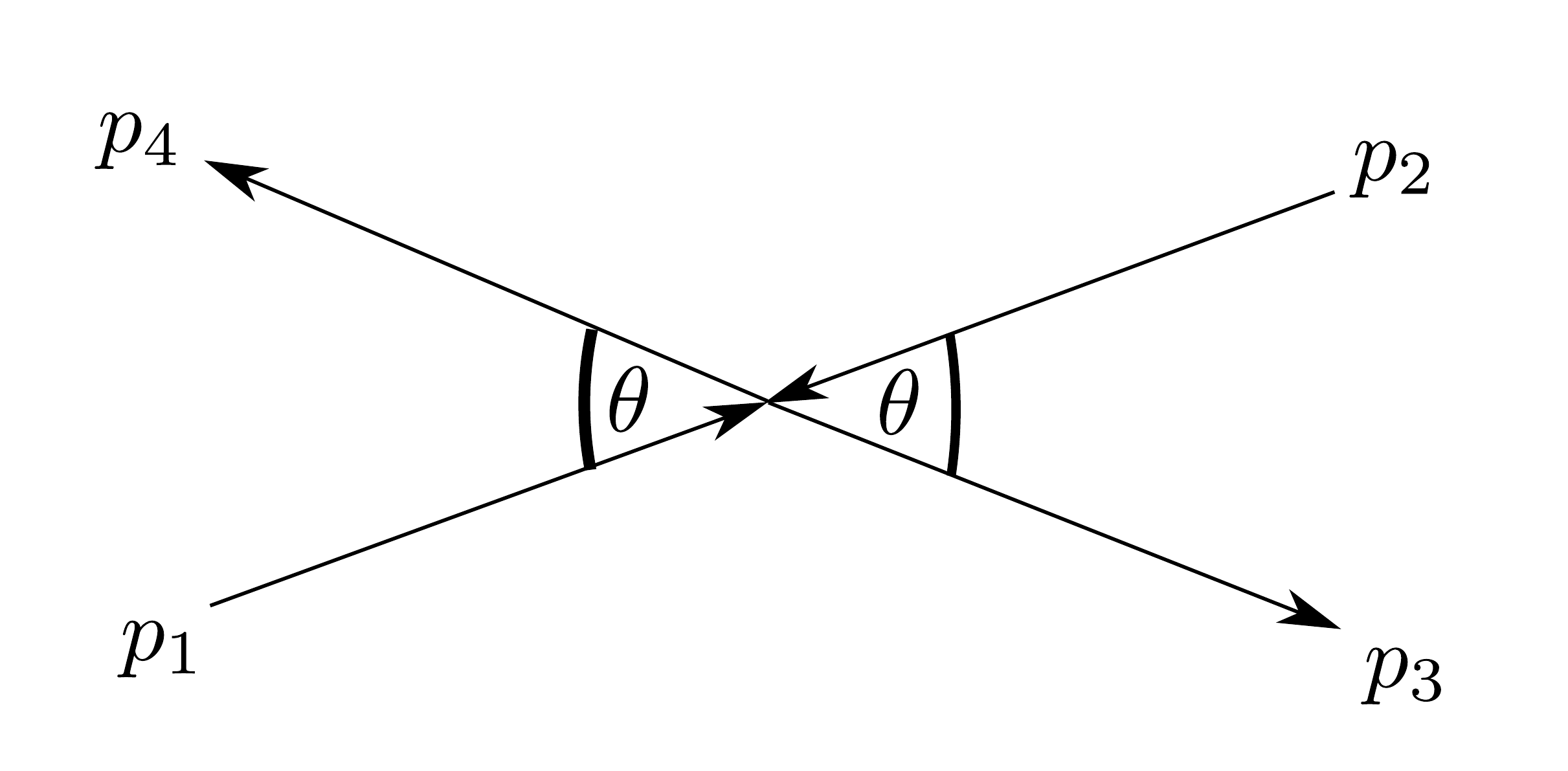}
\caption{\label{fig:int} Photon-photon scattering by the angle $\theta$ -- in-going momenta $p_1,\,p_2$ and out-going momenta $p_3,\,p_4$. The photons have the polarization $\xi_1,\,\xi_2$ and $\xi_3,\,\xi_4$ respectively, where $\xi_i\in \{1,2\}$.}
\end{figure}

Our observable, the differential cross section, can be derived from the scattering matrix $S$ \cite{Itzykson2006}. We define the scattering amplitude matrix $M$ such that 
\begin{eqnarray}
	\nonumber\langle p_3,\xi_3;p_4,\xi_4|S|p_1,\xi_1;p_2,\xi_2\rangle\\
	= \mathbb{I}+i\left(2\pi\right)^4\delta^{(4)}(p_1+p_2-p_3-p_4)M_{\xi_1\xi_2\xi_3\xi_4}(E,\theta)
\end{eqnarray}
The indices $\xi_i$ of the matrix elements $M_{\xi_1\xi_2\xi_3\xi_4}$ refer to the polarizations of the photons with momentum $p_1,\,p_2$ and $p_3,\,p_4$ in that order. The index $1$ refers to the linear polarization perpendicular to the plane of collision and the index $2$ refers to the linear polarization parallel to that plane.
In the center of momentum frame where $p_1=-p_2$ and $p_3=-p_4$, the components $M_{\xi_1\xi_2\xi_3\xi_4}$ are a function of only the energy $E$ and the scattering angle $\theta$ between $p_1$ and $p_3$ (see figure \ref{fig:int}) due to rotational symmetry of empty Minkowski space. For the sake of the simplicity of the expressions, we will not always write this dependence explicitly in the following.

The scattering amplitude and the differential cross section for the graviton mediated interaction of photons on the first tree-level were calculated in \cite{Barker1967erratum,Boccaletti1969}.
\begin{figure}
%
%
%
%
%
%
%
	\includegraphics[width=8cm,angle=0]{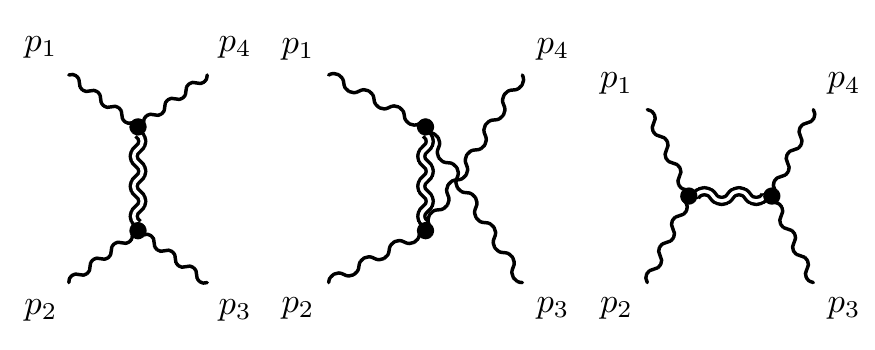}
	\caption[Feynman diagrams]{Contributing diagrams}
	\label{fig:compton-scattering}
\end{figure}
The nonzero elements of the scattering amplitude matrix for the in-going momenta $p_1\,,p_2$ and the out-going momenta $p_3,\,p_4$ (see Fig. \ref{fig:int}) in the center of momentum frame are found by evaluating the three contributing diagrams illustrated in Fig. \ref{fig:compton-scattering} (For the derivation, refer to the Appendix.). We obtain
\begin{eqnarray}\label{eq:amp}
\nonumber	M_{1111}&=& M_{2222}=\frac{\zeta^2 E^2 }{c^2\hbar\sin^2\theta}[-9-6\cos^2\theta-\,\,cos^4\theta]\\
	\nonumber M_{1122}&=& M_{2211}=\frac{\zeta^2 E^2 }{c^2\hbar\sin^2\theta}[\,\,\,\,7-6\cos^2\theta-\,\,cos^4\theta]\\
\nonumber	M_{1212}&=& M_{2121}=\frac{\zeta^2 E^2 }{c^2\hbar\sin^2\theta}[-8-4\cos\theta-4cos^3\theta]\\
\nonumber	M_{1221}&=& M_{2112}=\frac{\zeta^2 E^2 }{c^2\hbar\sin^2\theta}[-8+4\cos\theta+4cos^3\theta]\,,
\end{eqnarray}
The polarization averaged differential cross section is then given as (\cite{Itzykson2006})
\begin{eqnarray}\label{eq:diffcrossav}
	 \nonumber\frac{d\sigma}{d\Omega}&=&\frac{c^2\hbar^2}{64(2\pi)^2E^2}\frac{1}{4}\sum_{\xi_1, \xi_2, \xi_3, \xi_4=1}^2| M_{\xi_1 \xi_2 \xi_3 \xi_4}|^2\\
	&=&\frac{32\, l_P^4}{\lambda^2}\frac{\left[1+\cos^{16}\frac{\theta}{2}+\sin^{16}\frac{\theta}{2}\right]}{\sin^4\theta}\,,
\end{eqnarray}
where $l_P=\sqrt{\frac{G\hbar}{c^3}}\approx 1.6162 \times 10^{-35}m$ is the Planck length and $\lambda$ is the wavelength of the photons in the center of momentum frame. 

The values of the differential cross section (\ref{eq:diffcrossav}) are very small, e.g. for $\lambda=500\mathrm{nm}$ and for $\lambda=10\mathrm{nm}$, we find that the factor $\frac{32\, l_P^4}{\lambda^2}$ is of the order $10^{-126}\mathrm{m}^2$ and $10^{-123}\mathrm{m}^2$, respectively. This is extremely small even in comparison to the very small QED photon-photon scattering cross section \cite{DeTollis1965} which is of the order $10^{-72}\mathrm{m}^2$ and $10^{-62}\mathrm{m}^2$ for $\lambda=500\mathrm{nm}$ and $\lambda=10\mathrm{nm}$, respectively. And the QED photon-photon scattering is far from being directly observable in experiments. Thus, there is no chance to directly detect gravitational photon-photon scattering in the near future. However, it is of conceptual interest. 

In \cite{Westervelt1970,Schucker1990}, it was shown that the differential cross section in equation (\ref{eq:diffcrossav}) coincides with the classical differential cross section for the scattering of two light pulses for small scattering angles. In contrast to the classical differential cross section, however, equation (\ref{eq:diffcrossav}) does depend on the polarization of the photons. In the next section, we will show how this leads to a dependence of the differential cross section on the degree of polarization entanglement between the two photons.


\section{Entangled photons}

To investigate the effect of entanglement on the differential cross section, we consider the following parameterized state of two photons of momentum $p_1$ and $p_2$:
\begin{eqnarray}\label{eq:state}
 	|\Psi\rangle_{\varphi,\rho} &=& \cos\varphi|p_1,1;p_2,2\rangle + e^{i\rho}\sin\varphi |p_1,2;p_2,1\rangle\,,
\end{eqnarray}
where $\xi_1,\xi_2\in\{1,2\}$ in the state $|p_1,\xi_1;p_2,\xi_2\rangle$ label the polarization directions; $\xi_{i}=1$ refers the linear polarization perpendicular to the scattering plane and $\xi_{i}=2$ refers to the linear polarization in the scattering plane.
In equation (\ref{eq:state}), $\varphi\in[0,\pi/2]$ parameterizes the entanglement of the state; for $\varphi= 0$ and $\varphi=\pi/2$, the state is not entangled, and for $\varphi=\pi/4$, it is maximally entangled. The parameter $\rho\in[-\pi/2,3\pi/2)$ governs the relative phase of the superposed states $|p_1,1;p_2,2\rangle$ and $|p_1,2;p_2,1\rangle$. In particular, $|\Psi\rangle_{\pi/4,0}=|\Psi^+\rangle$ and $|\Psi\rangle_{\pi/4,\pi}=|\Psi^-\rangle$ are known as the symmetric and the anti-symmetric Bell states, respectively.

We are not interested in the final polarization state of the two photons. Therefore, we obtain the differential cross section (DCS) by summing over all final polarization states. Due to this summation, the DCS is independent of the choice of a polarization basis; we could have, equally well, written the state (\ref{eq:state}) in the basis of circular polarizations, without changing the result.

\begin{figure}
\includegraphics[width=8cm,angle=0]{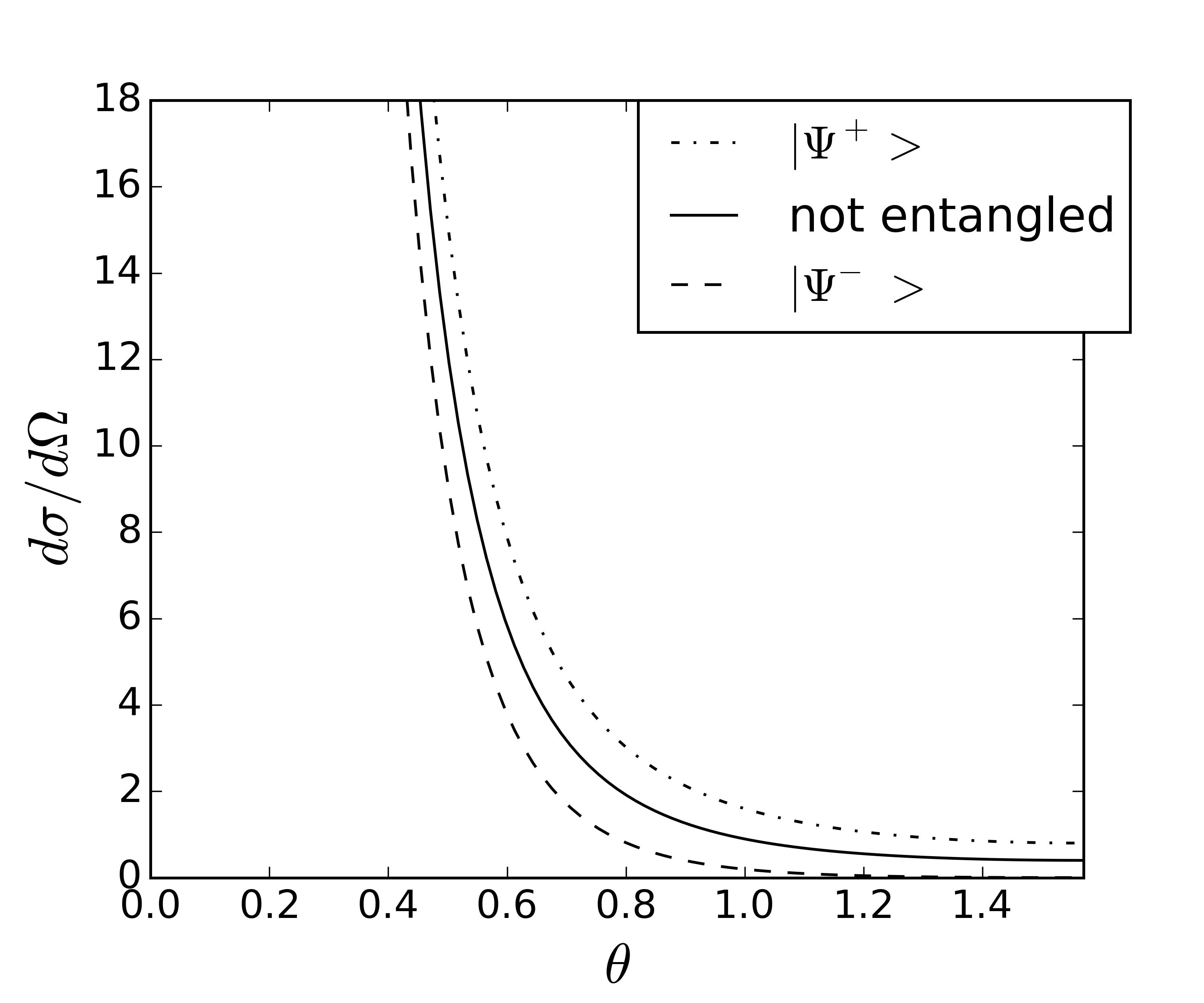}
\caption{\label{fig:plot_diff_pqg} The PQG differential cross section for not entangled photons and maximally entangled photons in the symmetric Bell state $|\Psi^+\rangle$ and the anti-symmetric Bell state $|\Psi^-\rangle$ in units of $8\frac{l_p^4}{\lambda^2}\times 10$.}
\end{figure}

At the first tree-level order, we obtain for the initial state (\ref{eq:state}) the DCS
\begin{eqnarray}\label{eq:diffcross}
	\frac{d\sigma_{|\Psi\rangle_ {\varphi,\rho}}}{d\Omega} &=& \frac{c^2\hbar^2}{64(2\pi)^2E^2}\times\\
	\nonumber &&\times  \sum_{\xi_3,\xi_4}
	 |\cos\varphi M_{12\xi_3\xi_4} + e^{i\rho}\sin\varphi M_{21\xi_3\xi_4}|^2\,.
\end{eqnarray}
Due to the rotational symmetry of Minkowski space, equation (\ref{eq:diffcross}) can be reduced to
\begin{eqnarray}\label{eq:diffcrossparity}
	\nonumber \frac{d\sigma_{|\Psi\rangle_ {\varphi,\rho}}}{d\Omega} = \frac{c^2\hbar^2}{64(2\pi)^2E^2} \left(|M_{1212}|^2+|M_{1221}|^2+\right.\\
	\left.+2\sin(2\varphi)\cos\rho\, \mathrm{Re}(M_{1212}M_{1221}^*)\right)\,.
\end{eqnarray}
Since the final particles are identical, we have $M_{\xi_1\xi_2\xi_3\xi_4}(E,\theta)=M_{\xi_1\xi_2\xi_4\xi_3}(E,\pi-\theta)$, and equation (\ref{eq:diffcrossparity}) can be expressed completely in terms of the matrix element $M_\theta:=M_{1212}$.
\begin{eqnarray}\label{eq:diffcrossparity_2}
	\nonumber \frac{d\sigma_{|\Psi\rangle_\varphi}}{d\Omega} &=&\nonumber  \frac{c^2\hbar^2}{64(2\pi)^2E^2} \left(|M_\theta|^2+|M_{\pi-\theta}|^2+\right.\\
	&& +2\sin(2\varphi)\cos\rho\cos\Delta\beta(\theta)\left.|M_\theta||M_{\pi-\theta}|\right)
\end{eqnarray}
where $\Delta\beta(\theta)$ is the relative phase of the complex functions $M_\theta$ and $M_{\pi-\theta}$. Using the matrix element in (\ref{eq:amp}) we obtain the differential cross section as
\begin{eqnarray}\label{eq:diffcross2}
	\frac{d\sigma_{|\psi\rangle_e}}{d\Omega}&=&\frac{8}{\sin^4\theta} \frac{l_P^4}{\lambda^2}\left[4(1+\sin(2\varphi)\cos\rho)+\right.\\
	&&\left.\nonumber +(1-\sin(2\varphi)\cos\rho)\left(cos\theta+\cos^3\theta\right)^2\right]\,,
\end{eqnarray}
where $l_P=\sqrt{\frac{G\hbar}{c^3}}\approx 1.6162 \times 10^{-35}m$ is the Planck length and $\lambda=\hbar c/E$ is the wavelength of the photons in the center of momentum frame. 
We find that the DCS in equation (\ref{eq:diffcross2}) is larger the stronger the entanglement if $-\pi/2<\rho<\pi/2$ and smaller the stronger the entanglement if $\pi/2<\rho<3\pi/2$. In particular, the strength of the interaction of the two photons reaches its maximum for the symmetric Bell state $|\Psi^+\rangle$ and its minimum for the anti-symmetric Bell state $|\Psi^-\rangle$. For $|\Psi^+\rangle$, $|\Psi^-\rangle$ and not entangled photons, the differential cross section is plotted in Fig. \ref{fig:plot_diff_pqg}). We find that for small scattering angles the DCS is independent of the entanglement parameter $\varphi$, and the effect of entanglement becomes significant only for large scattering angles. 
The effect of entanglement is maximal for right angle scattering, i.e. the scattering angle $\theta=\pi/2$. The DCS then becomes
\begin{eqnarray}\label{eq:diffcrossparity_0}
	\frac{d\sigma_{|\psi\rangle_e}}{d\Omega}&=&\frac{32}{\sin^4\theta} \frac{l_P^4}{\lambda^2}\left[1+\sin(2\varphi)\cos\rho\right]\,.
\end{eqnarray}
The DCS for right angle scattering vanishes for $|\Psi^-\rangle$ and is larger by a factor two for $|\Psi^+\rangle$ than for not entangled states.

\section{Comparison with QED}

In this section, we compare the effect of polarization entanglement in photon-photon scattering of PQG with the effect in Quantum Electrodynamics (QED). The interaction between photons in QED is the result of virtual electrons and positrons, created from the vacuum by the photons. In \cite{Karplus1950}, the following scattering amplitude for photon-photon scattering for QED was derived in the low energy limit (see also \cite{DeTollis1965,Liang2012}):
\begin{eqnarray}\label{eq:Mqed}
	iM_{1212} &=& \frac{4\alpha^2 E^4}{45m^4c^8}\left(31 + 22\cos\theta + 3\cos^2\theta\right)\,,
\end{eqnarray}
where $m$ is the electron mass and $\alpha$ is the fine structure constant. In contrast to PQG, the photon-photon scattering amplitude of QED in the low energy limit is dominated by virtual exchange particles (virtual electron-positron pairs) and the scattering amplitude is purely imaginary. However, equation (\ref{eq:diffcrossparity_2}) tells us that the dependence of the DCS on the entanglement is the same as in PQG if the complex phase of $M_\theta$ is independent of the scattering angle and there are no scattering angles $\theta<\pi$ for which $M_\theta$ vanishes. Hence, in QED - as in PQG -, the strength of the interaction of the two photons reaches its maximum for the symmetric Bell state $|\Psi^+\rangle$ and its minimum for the anti-symmetric Bell state $|\Psi^-\rangle$. 
The differential cross section follows from equation (\ref{eq:diffcrossparity_2}) as 
\begin{eqnarray}\label{eq:diffcrossqed}
	\frac{d\sigma_{|\psi\rangle_e}}{d\Omega}&=&\frac{\alpha^4}{2\cdot 45^2(2\pi)^2}\frac{\lambda_e^8}{\lambda^6}\times\\
	\nonumber &&\times\left[(1+\sin(2\varphi)\cos\rho)(31+3\cos^2\theta)^2+\right.\\
	\nonumber &&\left.\quad +(1-\sin(2\varphi)\cos\rho)22^2\cos^2\theta\right]\,,
\end{eqnarray}
where $\lambda_e=\hbar/mc$ is the Compton wavelength of the electron and $\lambda=\hbar c/E$ is the wavelength of the two photons. We see that the effect of polarization entanglement in QED is already significant for small scattering angles. This is because $M_{1221}(E,\theta)=M_{1212}(E,\pi-\theta)$ is non-zero for $\theta=0$ in QED; there is a non-zero amplitude for the process that can be interpreted as back scattering or polarization swapping. The dependence of the differential cross section on the parameters of $|\Psi\rangle_{\varphi,\rho}$ is the same in PQG and QED. This suggests that this dependence may be a general feature in photon-photon scattering. We will discuss this in the next section.

\section{Interpretation}

In this section, we give an interpretation of our results, firstly, in terms of quantum interference and, secondly, in terms of localized particles. In (\ref{eq:diffcross}), it can be seen that the dependence of the differential cross section on $\rho$ and $\varphi$, is an effect of quantum interference. The amplitude $M_{12\xi_3\xi_4}$ represents the scattering process in which the initial state was $|p_1,1;p_2;2\rangle$ and the final state is $|p_3,\xi_3;p_4;\xi_4\rangle$. The amplitude   $M_{21\xi_3\xi_4}$ represents the scattering process in which the initial state was $|p_1,2;p_2;1\rangle$ and the final state is $|p_3,\xi_3;p_4;\xi_4\rangle$. These amplitudes interfere to give rise to the scattering amplitude for the state $|\Psi\rangle_{\varphi,\rho}$. This situation is similar to the quantum interference of photons at a beam splitter known from the Hong-Ou-Mandel effect \cite{Hong1987}; photons in the symmetric Bell state interfere at a beam splitter such that they both always leave it on the same side.

The different resulting strengths of gravitational interaction for photons in the symmetric and anti-symmetric Bell states can also be interpreted in the sense of distance dependent forces. For this interpretation, we investigate the probability for the delayed coincidence measurement of two photons \cite{Glauber1963} 
\begin{eqnarray}
	P_{i\rightarrow f}\propto \sum_f \left|\langle f| E_l^{(+)}(t,x)E_j^{(+)}(t',x')|i\rangle\right|^2\,,
\end{eqnarray}
where $|f\rangle$ and $|i\rangle$ are the final and the initial state, respectively, and $E_k^{(+)}$ is the positive frequency part of the the $k$-component of the electric field operator
\begin{equation}
	E^{(+)}_k(t,x)=\int d\tilde{p}\, i\omega \,a_{p,k}e^{-\frac{i}{\hbar}p\cdot x + i\omega t}
	 \,.
\end{equation} 
For the initial state $|\Psi\rangle_{\varphi,\rho}$ in (\ref{eq:state}), we find for $l=2$ and $j=1$ 
\begin{eqnarray}\label{eq:coincidence}
	\nonumber &&\sum_f \left|\langle f| E_2^{(+)}(t,x)E_1^{(+)}(t',x')|\Psi\rangle_{\varphi,\rho}\right|^2\\
	 &\propto&\left[1+\sin(2\varphi)\cos\left(\frac{2}{\hbar}p\cdot(x'-x)+\rho\right)\right]\,.
\end{eqnarray}
This shows that the probability to find two photons at two given points closer than $\pi\lambda/4=\pi\hbar c/4E$ is increased for photons in the symmetric Bell state, $\rho=0$ and $\varphi=\pi/4$, and decreased for photons in the anti-symmetric Bell state, $\rho=\pi$ and $\varphi=\pi/4$, when compared to not entangled photons, $\varphi=0$ and $\varphi=\pi/2$. There is no difference to the not entangled case for the parameter values $\rho=\pm\pi/2$ independently of the parameter $\varphi$.

Hence, the dependence of the differential cross section on the degree of entanglement and the phase $\rho$ in $|\Psi\rangle_{\varphi,\rho}$ fits naturally with the idea of localized particles that interact via forces that decay with the distance between the particles, such as gravity and the QED photon-photon interaction. The photons in the symmetric Bell state are much more likely to be found at distances smaller than $\pi\lambda/4$ than not entangled photons which, in turn, are much more likely to be found at distances smaller than $\pi\lambda/4$ than photons in the anti-symmetric Bell state. Hence, photons in the symmetric Bell state interact more than not entangled photons which, again, interact more than photons in the anti-symmetric Bell state.

\section{Conclusions}

In the framework of Perturbative Quantum Gravity (PQG), the differential cross section for the scattering of two photons was derived  in \cite{Barker1967} and \cite{Boccaletti1969}. It was already noted in \cite{Barker1967} that the dependence of the gravitational interaction between photons on their polarization is in conflict with the weak equivalence principle.

We used the results of \cite{Boccaletti1969} and \cite{Barker1967} to show that polarization entangled photons gravitate more in the symmetric Bell state and less in the anti-symmetric Bell state. We compared the differential cross sections for photon-photon scattering in PQG with that in Quantum Electrodynamics. We found that they show the same dependence on the entanglement. We interpreted the results in the sense of quantum interference and in the sense of localized particles. We found that our results fit naturally into the idea of particles interacting via forces that decay with the distance between these particles. To work this out in more detail, it may be worthwhile to apply our approach to the interaction of photon wave packets. 

It would be interesting to find how these results relate to those of \cite{Bruschi2016}, how entanglement affects the gravitational self interaction of single particles in superposition states of different localizations. The maximally entangled state considered in \cite{Bruschi2016} is a symmetric state. In \cite{Bruschi2016}, the framework of semi-classical gravity was used with all its inherent conceptual problems (for more information on the conceptual problems of semi-classical gravity see \cite{Ford1982,Kuo1993,Phillips2000,Hu2000,Ford2003}). Even when ignoring these conceptual problems, it is not possible to derive the results presented here with semi-classical gravity. The effect we found here relies on the dependence of the gravitational interaction of photons on their polarization direction, but the classical gravitational field of light is independent of its polarization direction \cite{Raetzel2016}.

In a gravity theory with torsion such as the Einstein-Cartan theory and the Poincar\'e-Gauge theory of gravity \cite{Hehl1976} the gravitational field of light depends on the polarization direction. However, in these theories, the electromagnetic field cannot be coupled minimally to the gravitational field without loosing its gauge invariance \cite{Itin2003}. One must resort to non-minimal coupling, which leads to a modification of the constitutive tensor. One way to deal with such modifications was developed in \cite{Raetzel2011} and \cite{Giesel2012}. This framework could be used to investigate the polarization dependent gravitational interaction of photons in a semi-classical approach in order to compare the results with the findings that are presented here. 

For the sake of conceptual rigor, it would also be interesting to consider the gravitational interaction of entangled photons in the general boundary formulation of quantum field theory \cite{Oeckl2006,Banisch2013}, where the scattering process can be restricted to a compact spacetime region. It would also be of great interest to investigate the gravitational effect of entanglement in a background independent framework of quantum gravity, such as Loop Quantum Gravity.

\acknowledgments
DR thanks David Edward Bruschi, Friedrich W. Hehl and Philip C. E. Stamp for helpful discussions and Kiri Mochrie for proof reading the manuscript and providing writing assistance..

\section{Appendix: The photon-photon scattering amplitude in PQG}

Here we will show how the amplitude of photon-photon scattering can be derived using the Feynman rules of PQG (see \cite{Scadron2012}). Let us call the diagrams in Figure \ref{fig:compton-scattering} from the left to the right a), a') and b). Every photon-photon-graviton vertex in a), a') and b) contributes with a delta distribution to enforce momentum conservation and the following vertex factor (the expression in \cite{Scadron2012} must be multiplied by a factor 2 \cite{Ravndal2001,Nieves1998}):
\begin{eqnarray}
	\zeta T_{\mu\nu\beta\alpha}(p',p)&=&\zeta[{p'}_{\alpha}(p_{\mu}\eta_{\beta\nu}+p_{\nu}\eta_{\beta\mu})\\
	&&\nonumber +p_{\beta}({p'}_{\mu}\eta_{\alpha\nu}+{p'}_{\nu}\eta_{\alpha\mu})\\
	&&\nonumber -\eta_{\alpha\beta}({p'}_{\mu}p_{\nu}+p_{\mu}{p'}_{\nu})\\
	&&\nonumber +\eta_{\mu\nu}(p'\cdot p \,\eta_{\alpha\beta}-p_{\beta}{p'}_{\alpha}) \\
	&&\nonumber - p'\cdot p(\eta_{\mu\alpha}\eta_{\nu\beta}+\eta_{\mu\beta}\eta_{\nu\alpha})] \,,
\end{eqnarray}
where $p$ and $p'$ are the momenta of the two photons interacting at the vertex. The internal graviton line is associated with an integral over the graviton momentum $q$ and the graviton propagator which is given as
\begin{eqnarray}
	\frac{i\mathcal{P}_{\mu\nu\alpha\beta}}{q^2+i\epsilon}=\frac{i}{2(q^2+i\epsilon)}
	(\eta_{\mu\alpha}\eta_{\nu\beta}+\eta_{\mu\beta}\eta_{\nu\alpha}-\eta_{\mu\nu}\eta_{\alpha\beta})\,,
\end{eqnarray}
in the harmonic gauge.
The diagram a) leads to the expression
\begin{eqnarray}
	M_a&=&-\zeta^2 (\epsilon^\beta(p_3))^*\epsilon^\alpha(p_1)T_{\mu\nu\beta\alpha}(p_3,p_1)\times\\
	&&\nonumber \times\frac{\mathcal{P}^{\mu\nu\rho\sigma}}{(p_1-p_2)^2}(\epsilon^\delta(p_4))^*\epsilon^\gamma(p_1)T_{\rho\sigma\delta\gamma}(p_4,p_2)\,,
\end{eqnarray}
where $\epsilon(p_i)$ is the polarization vector corresponding to the photon $i$.
Adding up $M_a$, $M_{a'}$ and $M_b$, we arrive at the scattering amplitude $M$.

\bibliographystyle{ieeetr} 
\bibliography{entanggrav}

\end{document}